\documentclass[twoside]{article}
\usepackage{amssymb}

\usepackage{qic,epsfig}
\usepackage{amsmath}
\usepackage{graphics}
\usepackage{graphicx}
\usepackage{dcolumn}
\usepackage{bm}

\begin{document}


\setlength{\textheight}{8.0truein} 

\runninghead{ The Witness of Sudden Change of Geometric Quantum Correlation}  {Chang-shui Yu, Bo Li
and Heng Fan}

{\normalsize \textlineskip \thispagestyle{empty} \setcounter{page}{1} }

{\normalsize 
\copyrightheading{0}{0}{2003}{000--000} }

{\normalsize \vspace*{0.88truein} }

{\normalsize \alphfootnote
}

{\normalsize \fpage{1} }

{\normalsize 
\centerline{\bf
THE WITNESS OF SUDDEN CHANGE OF GEOMETRIC QUANTUM CORRELATION } \vspace*{0.035truein%
}  \vspace*{0.37truein} 
\centerline{\footnotesize
Chang-shui Yu$^1$\footnote{quaninformation@sina.com; ycs@dlut.edu.cn}, Bo Li$^2$, and Heng Fan$^{2}$} \vspace*{0.015truein} 
\centerline{\footnotesize\it
$^1$School of Physics and Optoelectronic Technology,}
\centerline{\footnotesize\it Dalian University of Technology, Dalian 116024, P. R. China }  
\centerline{\footnotesize\it $^2$Beijing National Laboratory for Condensed Matter Physics,}
\centerline{\footnotesize\it 
Institute of Physics, Chinese Academy of Sciences, Beijing 100190, China}\baselineskip=10pt \vspace*{10pt}
\vspace*{0.225truein} \publisher{(received
date)}{(revised date)} }

{\normalsize \vspace*{0.21truein} }

{\normalsize 
\abstracts{
In this paper, we give a sufficient and necessary condition (witness) for
the sudden change of geometric quantum discord by considering mathematical
definition of the discontinuity of a function. Based on the witness, we can
find out various sudden changes of quantum correlation by considering both
the Markovian and the non-Markovian cases. In particular, we can accurately
find out critical points of the sudden changes even though they are 
not quite obvious in the graphical representation. In addition, one can also find
that sudden change of quantum correlation, like the frozen quantum
correlation, strongly depends on the choice of the quantum correlation
measure. }{}{} }

{\normalsize \vspace*{10pt} }

{\normalsize \vspace*{3pt} \communicate{to be filled by the Editorial} }

{\normalsize \vspace*{1pt}\textlineskip    
}



\section{Introduction}

When quantum correlation is mentioned, entanglement could immediately come
to our head, because entanglement has attracted us so many years, in
particular it plays a very important role in quantum information processing
and is recognized as a necessary physical resource for quantum communication
and computation [1]. However, entanglement can not cover all the quantumness
of correlation in a quantum system, since some quantum tasks display the
quantum advantage without entanglement, for example, quantum discord, which
has been considered as the quantum correlation measure, is shown to be
possibly related to the speedup of some quantum computation [2-6]. It is
very interesting that quantum discord includes quantum entanglement but it
is beyond quantum entanglement due to its potential presence in separable
states [7].

In recent years, quantum discord has received great attention. A lot of
people studied its behavior under dynamical processes [8,9] and the
operational meanings by connecting it with Maxwell demon [10-12], or some
quantum information processes such as broadcasting of quantum states
[13,14], quantum state merging [15,16], quantum entanglement distillation
[17,18], entanglement of formation [19]. In particular, due to the
unavoidable interaction between the quantum system and the environment, it
has been found that quantum correlation in some cases [20,21] is more robust
against quantum decoherence than quantum entanglement [22,23]. Even the
frozen behavior of quantum correlation under some decoherence has been
reported [20,21,24,25]. However, like quantum entanglement measure, quantum
discord based on different definitions can lead to different results if we
use them to order two quantum states [26-28]. This implies that the behavior
of quantum correlation under coherence could depend on the choice of quantum
correlation measure and strong evidences have shown that the frozen behavior
will vanish if one employs a different quantum correlation measure [20]. In
addition, the sudden change of quantum correlation has also been found in
some dynamical processes (see [20-25] and the references therein). This
sudden change phenomenon is very important and physically useful, because it
is shown that the sudden change of some quantum systems is connected with
quantum phase transition (QPT) [28,29]. In particular, there are some
physical situations where entanglement is not able to detect QPT but this
sudden change are and even can be done at finite temperature. In addition,
unlike the sudden death of quantum entanglement, it seems that the sudden
change of quantum discord depends on the choice of the measure of quantum
correlation. Therefore, the interesting question that we would like to focus
is not only to find the models that demonstrate the behavior of sudden change of
quantum discord, but also to find out what in mathematics, leads to the sudden
change for a given quantum discord.

In this paper, we will study this question by considering a general quantum
systems with two qubits. Here we mainly employ the geometric quantum discord
[30] as the measure of quantum correlation due to its analytical
solvability. We give a mathematical definition of sudden change and
find a simple witness on the sudden change of geometric quantum discord
which also serves as a sufficient and necessary condition for the presence
of sudden change. Meanwhile, we find that the sudden change is only of one type. Based on our witness, we study various quantum systems
via the decoherence quantum channels and find many interesting phenomena of
sudden changes. For the usual quantum channel, we find that two critical
points of sudden change can be present compared with the previous similar
work. In particular, we can accurately find out the critical points no
matter whether the sudden change is obvious in the graphical representation.
For the non-Markovian case, one can find plenty of sudden changes so long as
we would like to properly adjust the corresponding parameters. By the state via the amplitude damping channels, we show that the 
critical points of sudden change with different quantum correlation measures appears at different positions. By the
collective decoherence, we demonstrate the inconsistence of sudden changes in both quantity and position
based on different correlation measures. This paper is organized as follows.
In Sec. II, we introduce the mathematical definition and the witness of
sudden change of quantum correlation. In Sec. III, we study various sudden
changes of geometric quantum discord compared with the information theoretic
quantum correlation in different quantum models. In Sec. IV, we draw the
conclusion.

\section{Witness on sudden change of geometric quantum discord}

In order to effectively understand the sudden change of quantum correlation,
we have to first give an explicit definition of the sudden change. Consider
a function $Q[\rho _{AB}(\xi )]$ serving as the measure of quantum
correlation, where it is implied that the quantum state $\rho _{AB}(\xi )$
depends on some parameter $\xi $, such as $\xi =\gamma t$ for decoherence
process with $\gamma $ the decoherence rate and $t$ the evolution time, we
say that $Q[\rho _{AB}(\xi )]$\textit{\ has sudden change or non-smooth at
some }$\xi ^{\ast }$\textit{, if }$\frac{dQ[\rho _{AB}(\xi )]}{d\xi }$%
\textit{\ is not continuous at }$\xi ^{\ast }$. On the contrary, we say $%
Q[\rho _{AB}(\xi )]$ is smooth if it has not any sudden changes, which
should be distinguished from the corresponding definition of a smooth
function in Mathematical Analysis that requires $Q[\rho _{AB}(\xi )]$ should
be of class $C^{\infty }$ [31].

Now we restrict our research to the process of decoherence. With
decoherence, the entries of the density matrix will exponentially decay in
general cases, therefore a reasonable hypothesis denoted by (H) is that the
evolution of the entries of the density matrix is smooth. In order to study
the interesting behavior of quantum correlation, we would like to employ the
analytic quantum correlation measure-----geometric quantum discord which is
defined for a general bipartite quantum state of qubits as 
\begin{equation}
D(\rho _{AB})=\frac{1}{4}(\left\Vert \vec{x}\vec{x}^{T}\right\Vert
^{2}+\left\Vert TT^{T}\right\Vert ^{2}-\lambda _{\max }),
\end{equation}%
where $\vec{x}=[x_{1},x_{2},x_{3}]^{T}$, $x_{i}=$Tr$\left[ \rho _{AB}\left(
\sigma _{i}\otimes \mathbf{1}\right) \right] $ with $\sigma _{i},i=1,2,3$,
corresponding to the three Pauli matrices, and $T_{ij}=$Tr$\left[ \rho
_{AB}\left( \sigma _{i}\otimes \sigma _{j}\right) \right] $ \ and $\lambda
_{\max }$ is the maximal eigenvalue of the matrix 
\begin{equation}
A=\vec{x}\vec{x}^{T}+TT^{T}.
\end{equation}%
In addition, $\mathbf{1}$ is the $(2\times 2)$ -dimensional identity and "$T$%
" in the superscript denotes the transpose of a matrix and $\left\Vert \cdot
\right\Vert $ is the Frobenius norm. With our hypothesis (H), it is obvious
that 
\begin{equation}
f(\rho _{AB})=\left\Vert \vec{x}\vec{x}^{T}\right\Vert ^{2}+\left\Vert
TT^{T}\right\Vert ^{2}
\end{equation}
is a simple function that directly depends on the entries of the density
matrix $\rho _{AB}$. So $f(\rho _{AB})$ is a smooth function. Thus the
sudden change of $D(\rho _{AB})$ has to attribute to $\lambda _{\max }$.
Therefore a quite simple and direct conclusion that can witness the sudden
change can be given in the following rigid way.

\textbf{Theorem 1.}-Sudden change will happen for geometric quantum discord
under decoherence if and only if $\lambda _{\max }$ is non-smooth.

In order to strengthen our understanding of the non-smooth behavior of
quantum correlation, we will expand the Theorem further. Consider the
eigenequation of $A$, we have [32] 
\begin{equation}
\lambda ^{3}+a_{2}\lambda ^{2}+a_{1}\lambda +a_{0}=0,
\end{equation}%
where $a_{0}=-\det A$, $a_{2}=-$Tr$A$ and $a_{1}=\sum_{k=1}^{3}\det  \mathcal A_{k}$
with $\mathcal A_{k}=\left( 
\begin{array}{cc}
A_{kk} & A_{k,k\oplus 1} \\ 
A_{k\oplus 1,k} & A_{k\oplus 1,k\oplus 1}%
\end{array}%
\right) $ and $"\oplus "$ denoting addition modulo 3, then the eigenvalues
of $A$ can be given by%
\begin{eqnarray}
\lambda _{1} &=&M_{+}^{1/3}+M_{-}^{1/3}-\frac{1}{3}a_{2},  \nonumber \\
\lambda _{2} &=&-\frac{M_{+}^{1/3}+M_{-}^{1/3}}{2}+i\sqrt{3}\frac{%
M_{+}^{1/3}-M_{-}^{1/3}}{2}-\frac{1}{3}a_{2}, \\
\lambda _{3} &=&-\frac{M_{+}^{1/3}+M_{-}^{1/3}}{2}-i\sqrt{3}\frac{%
M_{+}^{1/3}-M_{-}^{1/3}}{2}-\frac{1}{3}a_{2},  \nonumber
\end{eqnarray}%
where $M_{\pm }=-\frac{q}{2}\pm \sqrt{\Delta }$, $\ \Delta =\frac{q^{2}}{4}+%
\frac{p^{3}}{27}$ with $p=a_{1}-\frac{1}{3}a_{2}^{2}$ and $q=a_{0}-\frac{1}{3%
}a_{1}a_{2}+\frac{2}{27}a_{2}^{3}$. The derivative $\frac{d\lambda _{i}}{dt}$
can be formally written as 
\begin{equation}
\frac{d\lambda _{i}}{dt}=F(M^{-2/3},\Delta ^{-1/2},\cdots ),
\end{equation}%
where we omit the smooth parameters. So one could imagine that the discontinuity of $\frac{d\lambda
_{\max }}{dt}$ could happen when $M=0$ or $\Delta =0$ for the possible unbounded derivative. However, for an infinitesimal evolution of a density matrix, $A(\delta t)$ can always be understood as an infinitesimal symmetric perturbation $E\delta t$ on the original $A$. Thus based on the perturbation theory [33], $\left\vert \lambda_i(A(\delta t))-\lambda_i(A)\right\vert\leq\left\Vert E\right\Vert\delta t$, which guarantees that no unbounded derivative can occur. So all the derivatives of the eigenvalues are continuous.
Consider that the maximal eigenvalue is required for the geometric discord,
one will draw the conclusion that the sudden change will happen only the
following corollary holds.

\textbf{Corollary 1}.-Sudden change will happen at $t^{\ast }$, if any and
only if there exists an eigenvalue $\lambda _{i}(t^{\ast })$ such that $%
\lambda _{i}(t^{\ast })$ and $\lambda _{\max }(t^{\ast })$ are crossing.
That is, if $\lambda _{\max }(t^{\ast }-\varepsilon )=\lambda _{m}(t^{\ast
}-\varepsilon )$ $\ $and $\lambda _{\max }(t^{\ast }+\varepsilon )=\lambda
_{n}(t^{\ast }-\varepsilon )$ for any small $\varepsilon $, we have $m\neq
n. $

Here we should distinguish the word `crossing' from `degenerate'. Finally,
we would like to emphasize that the similar idea can also be used in other
type of quantum correlation measures, which is also demonstrated in our next
section.

\section{Various sudden changes of geometric quantum discord}

Using the witness proposed in previous section, we can easily find various
sudden changes of geometric quantum discord for some states separately
undergoing some decoherence channels. In particular, even though the
critical points of sudden change could not be obvious in the graphical
representation, we can also accurately find them out. Next, we consider the
sudden changes in both the Markovian and the non-Markovian cases.

\textit{Markovian case.}-Let's first consider the initial state given by 
\begin{equation}
\rho _{AB}=\frac{1}{4}\left[ \mathbf{1}_{AB}+\sum\limits_{i}(c_{i0}\sigma
_{i}\otimes \sigma _{i})\right] ,
\end{equation}%
where $\sigma _{i}$ defined the same as that in Eq. (1) and $\left\vert
c_{i0}\right\vert \leq 1$ with $\sum\limits_{i}\left\vert c_{i0}\right\vert
\leq 1$ and the subscript $0$ denotes the initial state. Suppose subsystem A
undergoes a phase damping quantum channel which is given in the Kraus
representation [34] as%
\begin{equation}
A_{1}=\sqrt{1-p/2}\mathbf{1},A_{2}=\sqrt{p/2}\sigma _{3},
\end{equation}%
and subsystem B goes through a bit flip quantum channel given by%
\begin{equation}
B_{1}=\sqrt{1-q/2}\mathbf{1},B_{2}=\sqrt{q/2}\sigma _{1},
\end{equation}%
where $p=1-e^{-\gamma _{1}t}$ and $\ q=1-e^{-\gamma _{2}t}$ with $\gamma
_{1,2}$ denoting decoherence rate, the evolution of $\rho _{AB}$ can,
therefore, be expressed as%
\begin{equation}
\$(\rho _{AB})=\sum\limits_{i,j=1}^{2}\left( A_{i}\otimes B_{j}\right) \rho
_{AB}\left( A_{i}^{\dag }\otimes B_{j}^{\dag }\right) .
\end{equation}%
In Bloch representation, $\$(\rho _{AB})$ can be written by the same form as
Eq. (7) with 
\begin{equation}
c_{1}=c_{10}e^{-\gamma _{1}t},c_{2}=c_{20}e^{-(\gamma _{1}+\gamma
_{2})t},c_{3}=c_{30}e^{-\gamma _{2}t}.
\end{equation}%
One should note that $c_{i}(t)$ serves as the eigenvalues of the matrix $A$
mentioned in Eq. (2) for $\$(\rho _{AB})$. Thus the geometric quantum
discord of $\$(\rho _{AB})$ can be easily calculated based on Eq. (1). One
can find that $c_{i}(t)$ is obviously an smooth function on time $t$, so the
critical point of sudden change is completely determined by the corollary. To demonstrate the sudden change, we would like to let $\left\vert
c_{20}\right\vert >\left\vert c_{10}\right\vert >\left\vert
c_{30}\right\vert >0$ and $\gamma _{1}<\gamma _{2}$. In this case, one can
find that $c_{1}$ and $c_{2}$ are crossing at $t_{1}=\frac{1}{\gamma _{2}}%
\ln \left\vert \frac{c_{20}}{c_{10}}\right\vert $, so a sudden change will
happen. In addition, we emphasize that no else sudden change but $t_{1}$ can
be found in the whole process of the evolution, which is implied in the case
of Ref. [20]. However, if we suppose $\left\vert c_{20}\right\vert
>\left\vert c_{10}\right\vert >\left\vert c_{30}\right\vert >0$ and $\gamma
_{1}>\gamma _{2}$, it is surprising that one can find two sudden changes in
the evolution. One happens at $t_{1}$ and the other happens at $t_{2}=\frac{1%
}{(\gamma _{1}-\gamma _{2})}\ln \left\vert \frac{c_{10}}{c_{20}}\right\vert $%
. In order to explicitly illustrate the sudden changes, we plot the
geometric quantum discord of $\$(\rho _{AB})$ in Fig. 1, where we let $%
c_{10}=0.12$, $c_{20}=0.13$, $c_{30}=0.08$ and $\gamma _{1}=0.035$, $\gamma
_{2}=0.015$. One can find that the figure is consistent to our prediction,
that is, the points of crossing eigenvalues of $A$ witness the sudden
changes. As a comparison, we also plot the information theoretic quantum
discord defined by the discrepancy between the quantum versions of the two
equivalent mutual information [2,4]. It is interesting that, for the state $%
\$(\rho _{AB})$, information theoretic quantum discord and the geometric
quantum discord have the same critical points for sudden change. 
\begin{figure}[htbp]
\centerline {\epsfig{file=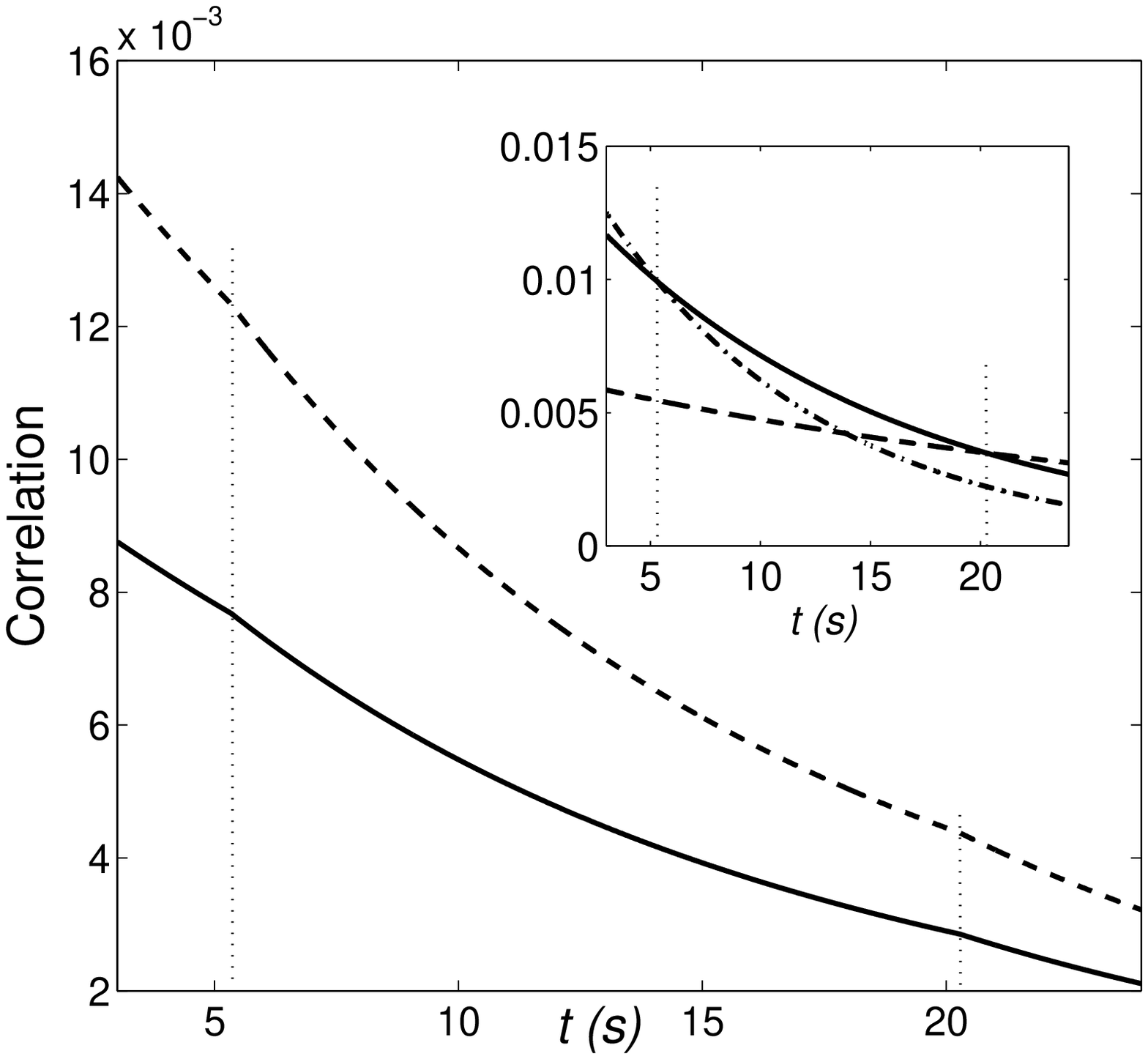,width=0.8\columnwidth}}
\vspace*{13pt}
\fcaption{\label{1}(Dimensionless) The doubled geometric quantum discord and
information theoretic quantum discord \textit{vs.} time $t$. The solid line
is information theoretic and dashed line is geometric. The inset shows the
evolution of the three eigenvalues of matrix $A$. All the vertical dotted
lines point out the critical points of sudden changes. The left critical
point is about $t=5.3362s$ and the right one is about $t=20.2733s$. }
\end{figure}

Now let's consider that the two qubits A and B of the state
given in Eq. (7) simultaneously undergo the phase damping channels which is
given by Eq. (8). That is, the state through the channels can be formally
given by Eq. (10) with $B_{j}$ replaced by $A_{j}$ and $\gamma _{1}$ replaced
by a new parameter $\gamma _{2}$. Thus the final state $\$(\rho _{AB})$ can
also be written as Eq. (7) with 
\begin{equation}
c_{1}=c_{10}e^{-(\gamma _{1}+\gamma _{2})t},c_{2}=c_{20}e^{-(\gamma
_{1}+\gamma _{2})t},c_{3}=c_{30}.
\end{equation}
We plot the geometric and the information theoretic quantum discords in Fig.
2, where we let $c_{10}=0.5$, $c_{20}=0.3$, $c_{30}=0.4$, $\gamma _{1}=0.45$
and $\gamma _{2}=0.15$. One can find that in the given range, there is only
one critical point of sudden change at $t=1/(\gamma_1+\gamma_2)\ln{\frac{5}{4}} s$ and this sudden change is
consistent with information theoretic quantum discord. 
\begin{figure}[htbp]
\centerline{ \epsfig{file=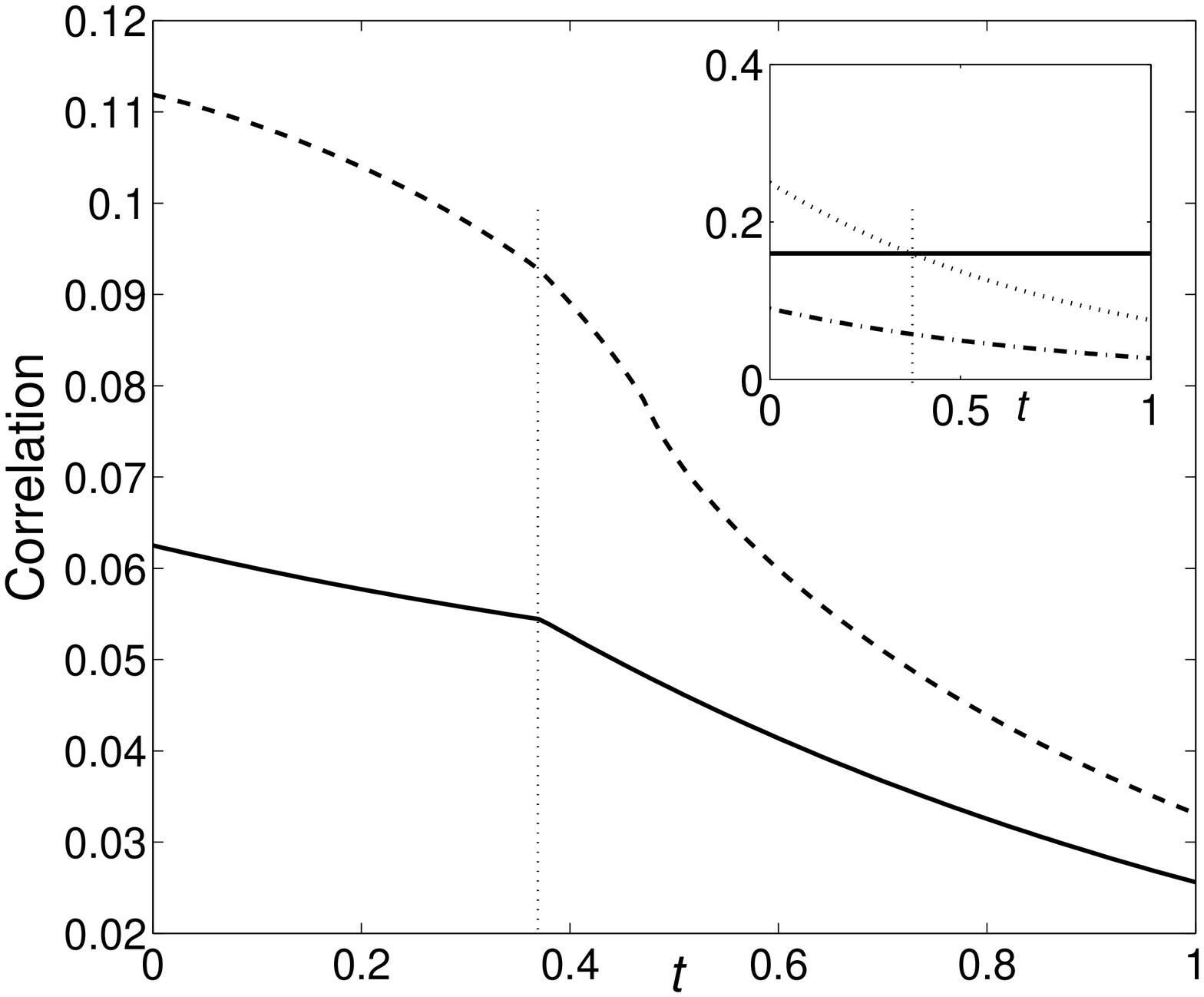, width=0.8\columnwidth}}
\vspace*{13pt}
\fcaption{\label{2}(Dimensionless) The geometric quantum discord and
the one-quarter information theoretic quantum discord \textit{vs.} time $t$. The dashed line
is information theoretic and solid is geometric. The
evolution of the three eigenvalues of matrix $A$ is given in the inset. The vertical dotted
line points out the critical point of sudden change at about $t=0.3719s$. }
\end{figure}

\textit{Non-Markovian case.}-The above example demonstrates the Markovian
process. Now we would like to consider the sudden change behavior in the
non-Markovian decoherence. Similar to the above case, we also consider the
two qubits separately undergo a single-direction quantum channel. We suppose
that subsystem A goes through a colored noise phase flip channel and
subsystem B undergoes a colored noise bit flip channel [21]. This can be
realized in Kraus representation by replacing $p$ and $q$ in Eqs. (10) and
(11) by%
\begin{equation}
x_{i}=1-e^{-\upsilon _{i}}\left[ \cos \left( \mu _{i}\upsilon _{i}\right)
+\sin \left( \mu _{i}\upsilon _{i}\right) /\mu _{i}\right] ,
\end{equation}%
where $\mu =\sqrt{(4a_{i}\tau _{i})^{2}-1}$, $a_{i}$ is a coin-flip random
variable and $\upsilon _{i}=t/(2\tau _{i})$ is dimensionless time with $i=1,2
$ corresponding to $p$ and $q$, respectively. Due to the smooth dependence
on the dimensionless time $\upsilon _{i}$, one can draw the conclusion that
the sudden change of geometric quantum discord is determined by the corollary. We plot the geometric quantum discord of the state $\rho _{AB}$
through two non-Markov quantum channels in Fig. 4, where we set $\tau
_{1}=\tau _{2}=5s$ and $a_{1}=2/3$, $a_{2}=1/3$. One can find more sudden
changes in this case. It is interesting that we can find various sudden
changes as we will in this case, because we can adjust the parameters based
on our witness to produce various crossing points of the eigenvalues of $A$.
It is obvious that the critical points of sudden changes of geometric
quantum discord are the same as those of information theoretic quantum
discord which is also plotted in Fig. 4. as a comparison.

\textit{Inconsistence of sudden changes with different measures.}-Since we know that it is of strong
dependence on the selected quantum correlation measure for frozen quantum
discord, that is, even though we have found the phenomenon of frozen quantum
discord for some measure, the frozen phenomenon will vanish if we change
into another quantum correlation measure, this means that the frozen
phenomenon is not a property of the quantum state, but some property of the
selected quantum correlation measure subject to some states. However, from
the above examples, one can find that the critical points are the same
between the two selected quantum correlation measure. One could ask a very
natural question whether the critical points of the sudden change are
independent of quantum correlation measure just as the sudden death of
quantum entanglement does not depend on the entanglement measure [35,36]. To
answer this question, let's consider two examples separately in the decoherence of amplitude damping [37] and the collective decoherence
[38-40].

At first, let's consider the case  of the amplitude damping decoherence. Suppose the
two qubit A and B of the state in Eq. (7) through the amplitude damping
\begin{figure}[htbp]
\centerline{ \epsfig{file=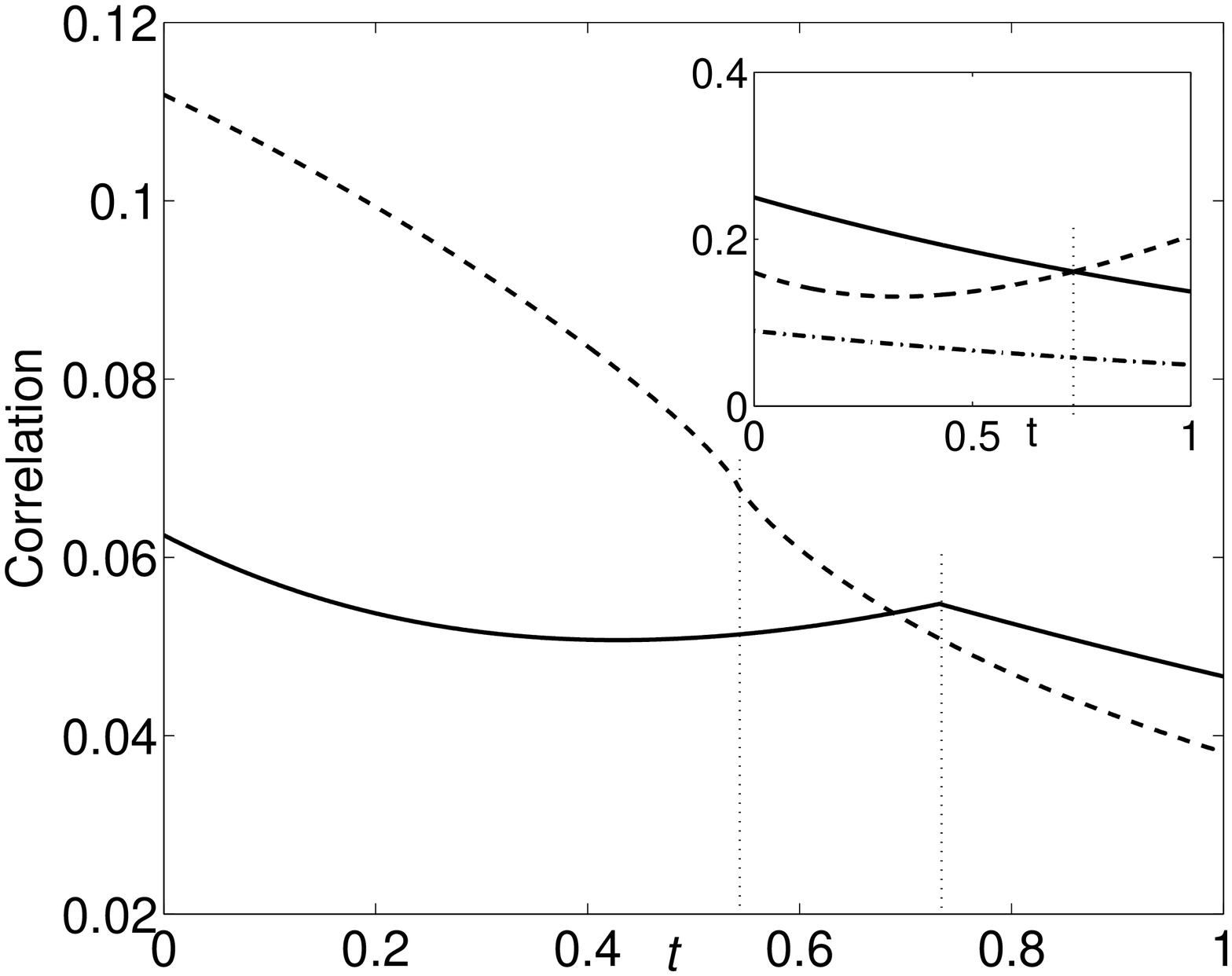,width=0.8\columnwidth}}
\vspace*{13pt}
\fcaption{\label{3}(Dimensionless) The geometric quantum discord and one-quarter
information theoretic quantum discord \textit{vs.} time $t$. The dashed line
is information theoretic and solid line is geometric. The inset shows the
evolution of the three eigenvalues of matrix $A$. The left vertical dotted
line points out the sudden change of the information theoretic discord and the right one points
out the sudden change of the geometric discord. }
\end{figure}channel, respectively. The amplitude damping channel in the Kraus
representation is written as [34]%
\begin{equation}
\tilde{A}_{k1}=\left( 
\begin{array}{cc}
1 & 0 \\ 
0 & \sqrt{1-p_{k}}%
\end{array}%
\right) ,\tilde{A}_{k2}=\left( 
\begin{array}{cc}
0 & \sqrt{p_{k}} \\ 
0 & 0%
\end{array}%
\right) ,
\end{equation}%
with $k=A,B$ corresponding to the two subsystems and $p_{k}=1-e^{-\gamma
_{k}t}$. However, unlike the previous case, the final state via the channels
will become a general ``X" type state instead of the state
given in Eq. (7). Thus even though we can analytically calculate the
geometric discord, one can only find the numerical expression for the
information theoretic discord. Both discords are plotted in Fig. 3 from
which one can find that the geometric discord has one sudden change at $%
t=0.732s$, but the information theoretic discord has one sudden change at
about $t=0.542s$. This example shows the critical points of the information theoretic and the geometric discords are at different positions.

Our system in our second example includes two identical atoms with $\left\vert
g\right\rangle $ and $\left\vert e\right\rangle $ denoting the ground state
and the excited state and $\varpi $ supposed to the transition frequency. We
assume the two atoms are coupled to a multimode vacuum electromagnetic
field. The master equation governing the evolution is given by%
\begin{eqnarray}
\dot{\rho} =-i\varpi \sum_{i=1}^{2}\left[ \sigma _{i}^{z},\rho \right]
-i\sum_{i\neq j}^{2}\Omega _{ij}[\sigma _{i}^{+}\sigma _{j}^{-},\rho ] +%
\frac{1}{2}\sum_{i,j=1}^{2}\gamma _{ij}\left( 2\sigma _{j}^{-}\rho \sigma
_{i}^{+}-\left\{ \sigma _{i}^{+}\sigma _{j}^{-},\rho \right\} \right) ,
\end{eqnarray}
where 
\begin{equation}
\gamma _{ij}=\frac{3}{2}\gamma \left[ \frac{\sin (kr_{ij})}{kr_{ij}}+\frac{%
\cos (kr_{ij})}{\left( kr_{ij}\right) ^{2}}-\frac{\sin (kr_{ij})}{\left(
kr_{ij}\right) ^{3}}\right]
\end{equation}
denotes the collective damping with $\gamma $ the spontaneous emission rate
due to the interaction between one atom and its own environment; 
\begin{equation}
\Omega _{ij}=\frac{3}{4}\gamma \left[ \frac{\sin (kr_{ij})}{\left(
kr_{ij}\right) ^{2}}+\frac{\cos (kr_{ij})}{\left( kr_{ij}\right) ^{3}}-\frac{%
\cos (kr_{ij})}{kr_{ij}}\right]
\end{equation}%
represents the dipole-diple interaction potential with $r_{ij}=\left\vert 
\mathbf{r}_{i}-\mathbf{r}_{j}\right\vert $ being the interatomic distance.
In addition, $k$ in Eqs. (16) and (17) is the wave vector. 
\begin{figure}[htbp]
\centerline{ \epsfig{file=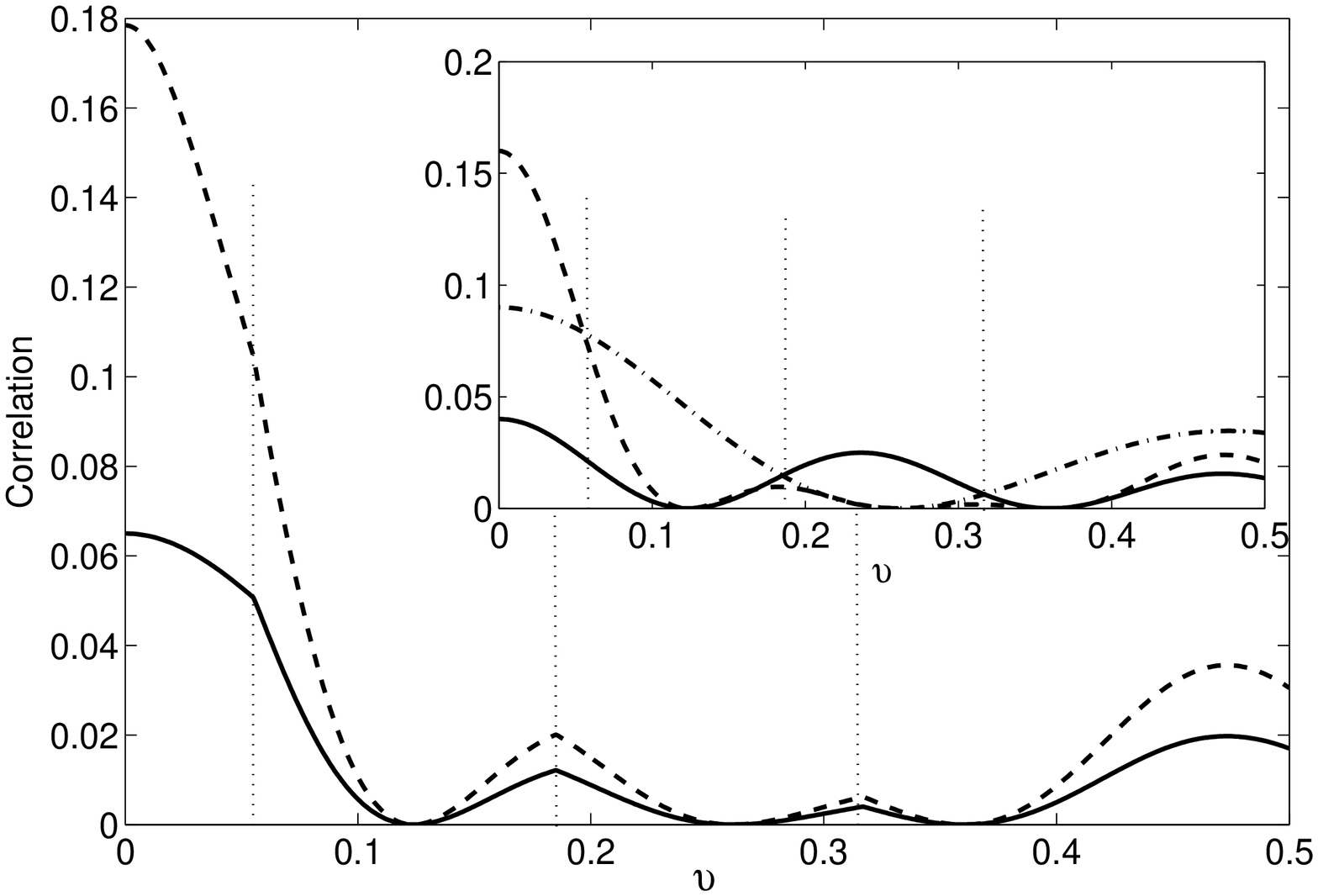,width=0.8\columnwidth}}
\vspace*{13pt}
\fcaption{\label{4}(Dimensionless) Quantum correlation vs. dimensionless time $\protect%
\upsilon$. The dashed line corresponds to the information theoretic quantum
discord and the solid line denotes to the geometric one. The inset shows the
evolution of the three eigenvalues of matrix $A$ and the signed crossing
points are the critical points of sudden change. The three sudden changes
happen at about $\protect\upsilon=0.055, \protect\upsilon= 0.185$ and $%
\protect\upsilon=0.317$.}
\end{figure}

Suppose the initial state of the two atoms is given by $\rho _{AB}(t)=[\rho _{ij}(t)]$ with 
\begin{eqnarray}
\rho _{11}(t) &=&\alpha ^{2}e^{-2\gamma t}, 
\end{eqnarray}
\begin{eqnarray}
\rho _{14}(t) &=&\rho _{41}^{\ast }(t)=\alpha \sqrt{1-\alpha ^{2}}e^{-\left(
\gamma +2i\varpi \right) t}, 
\end{eqnarray}
\begin{eqnarray}
\rho _{22}(t) &=&\rho _{33}(t)=a_{1}\left[ e^{-\gamma _{12}^{+}t}-e^{-\gamma
t}\right] +a_{2}[e^{-\gamma _{12}^{-}t}-e^{-\gamma t}], 
\end{eqnarray}
\begin{eqnarray}
\rho _{23}(t) &=&\rho _{32}(t)=a_{1}\left[ e^{-\gamma _{12}^{+}t}-e^{-\gamma
t}\right] -a_{2}[e^{-\gamma _{12}^{-}t}-e^{-\gamma t}], 
\end{eqnarray}
\begin{eqnarray}
\rho _{44} &=&1-\rho _{11}(t)-\rho _{22}(t)-\rho _{33}(t).
\end{eqnarray}%
where $\gamma _{12}^{\pm }=\gamma \pm \gamma _{12}$ and $a_{1,2}=\alpha
^{2}\gamma _{12}^{\pm }/\left( 2\gamma _{12}^{\mp }\right) $. If we
calculate $f(\rho _{AB}(t))$ defined by Eq. (3), we can find that $f$
smoothly depends on $t$, so we will focus on the matrix $A$ defined by Eq.
(2). Under some local unitary transformation, we can find that the
eigenvalues of the matrix $A$ can be given by%
\begin{eqnarray}
\lambda _{\pm }(A) &=&4(\rho _{23}\mp \left\vert \rho _{14}\right\vert )^{2},
\\
\lambda _{0}(A) &=&C^{2}+R^{2},
\end{eqnarray}%
with $C=1-4\rho _{22}$ and $R=2\rho _{11}+2\rho _{22}-1$. Thus we can
directly check where the eigenvalues are crossing or non-smooth in order to
find out the sudden change, since the eigenvalues given in Eqs. (5-7) are
obviously smoothly dependent of time $t$. The evolution of the eigenvalues
and the geometric quantum discord is plotted in Fig. 6 and Fig. 5, respectively, where we set $\alpha =%
\sqrt{0.9}$ and $r_{12}=0.6737\lambda $ with $\lambda $ the wave length. 
 From
the Fig. 5. and Fig. 6, one can find that the critical points 
of sudden change are consistent with the crossing points of the maximal
eigenvalues. In order to compare it with information theoretic quantum
discord, we would like to give the explicit expression of the quantum
discord as 
\begin{equation}
D^{\prime }(\rho _{AB})=1+H(R)+\min_{i=0,\pm }\left\{ s_{i}\right\}
-\sum\limits_{i=\pm }\left( u_{i}\log _{2}u_{i}+v_{i}\log _{2}v_{i}\right) ,
\end{equation}
where 
\begin{equation}
H(x)=-(1-x)\log _{2}(1-x)-(1+x)\log _{2}(1+x),
\end{equation}%
and%
\begin{eqnarray}
u_{\pm } &=&\frac{1}{4}(1-C\pm 4\rho _{23}), \\
v_{\pm } &=&\frac{1}{4}(1+C\pm 2\sqrt{R^{2}+\left\vert \rho _{14}\right\vert
^{2}}),
\end{eqnarray}%
and 
\begin{eqnarray}
s_{\pm } &=&1+H(\sqrt{R^{2}+\lambda _{\pm }}), \\
s_{0} &=&-\sum\limits_{i=\pm }\left[ \frac{m_{i}}{4}\left( \log _{2}\frac{%
m_{i}}{n_{i}}\right) +\frac{1-C}{4}\log _{2}\frac{1-C}{n_{i}}\right] ,
\end{eqnarray}%
with 
\begin{eqnarray}
m_{\pm } &=&\frac{1+C\pm 2R}{4}, \\
n_{\pm } &=&2\left( 1\pm R\right) .
\end{eqnarray}%
\begin{equation}
\left\vert \Psi \right\rangle _{AB}=\alpha \left\vert e\right\rangle
_{A}\left\vert e\right\rangle _{B}+\sqrt{1-\alpha ^{2}}\left\vert
g\right\rangle _{A}\left\vert g\right\rangle _{B},
\end{equation}%
then the state via the evolution subject to Eq. (15) at time $t$ can be
\begin{figure}[htbp]
\centerline{\epsfig{file=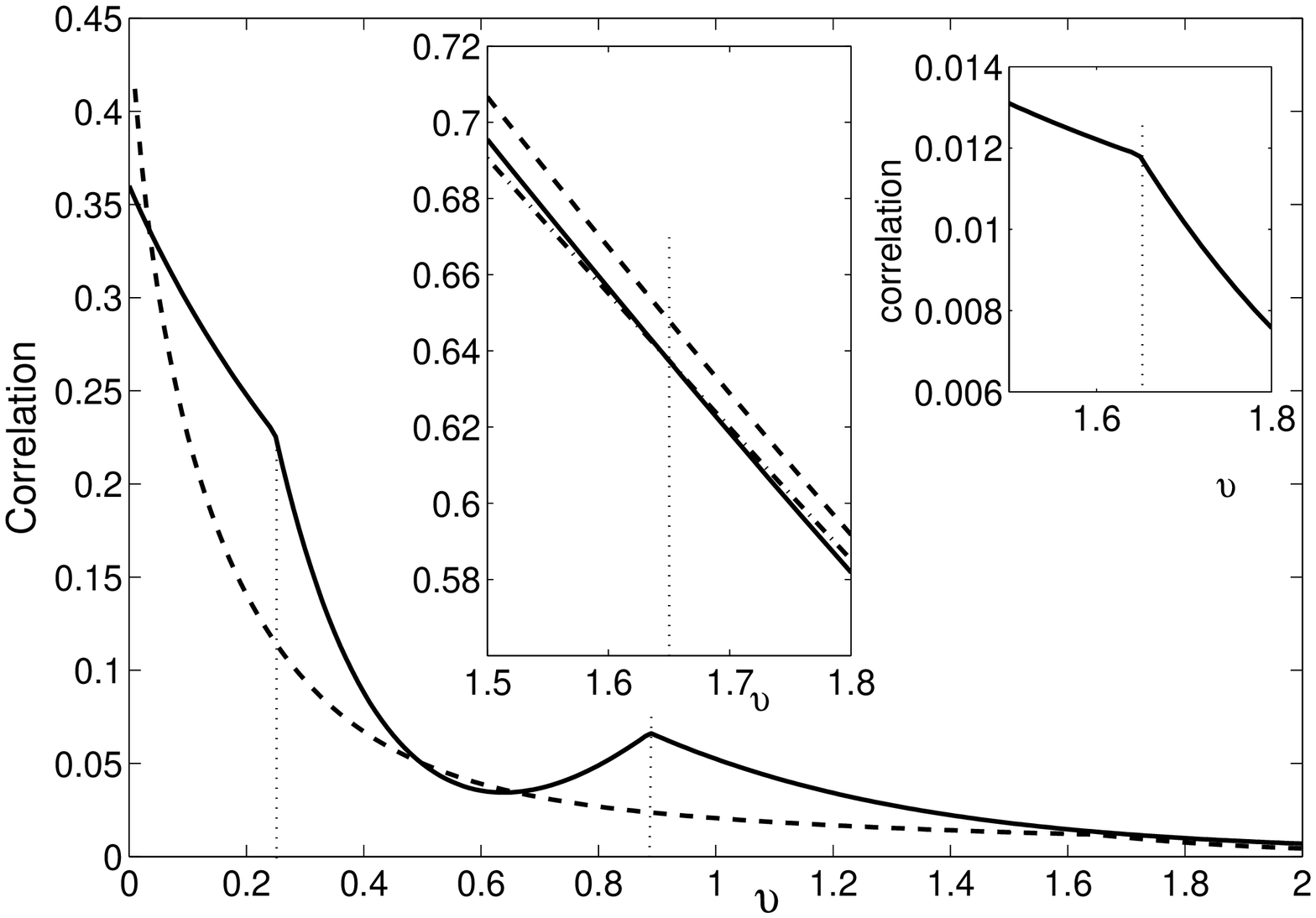,width=0.8\columnwidth}}
\vspace*{13pt}
\fcaption{\label{5}(Dimensionless) Quantum correlation vs. $\protect\upsilon =\protect%
\gamma t$. The solid line is the doubled geometric quantum discord and the
dashed line is the information theoretic one. The left inset shows the
evolution of $\left\vert c_{i}\right\vert $ and the crossing point of the
two smaller one is the critical point of sudden change for information
theoretic quantum discord. The right inset shows the enlarged unique
critical point at about $\protect\upsilon =1.65$ for the information
theoretic quantum discord. The two critical points for geometric quantum
discord are at about $\protect\upsilon =0.25$ and $\protect\upsilon =0.89$,
respectively.}
\end{figure}
From Eq. (26), we can draw the conclusion that the sudden changes are only
determined by the crossing of $s_{i}$, because all the other terms in Eq.
(26) are smooth on $t$ which can be found by continuous derivatives of these
terms. Therefore, we can find the critical points of sudden changes for
\begin{figure}[htbp]
\centerline{ \epsfig{file=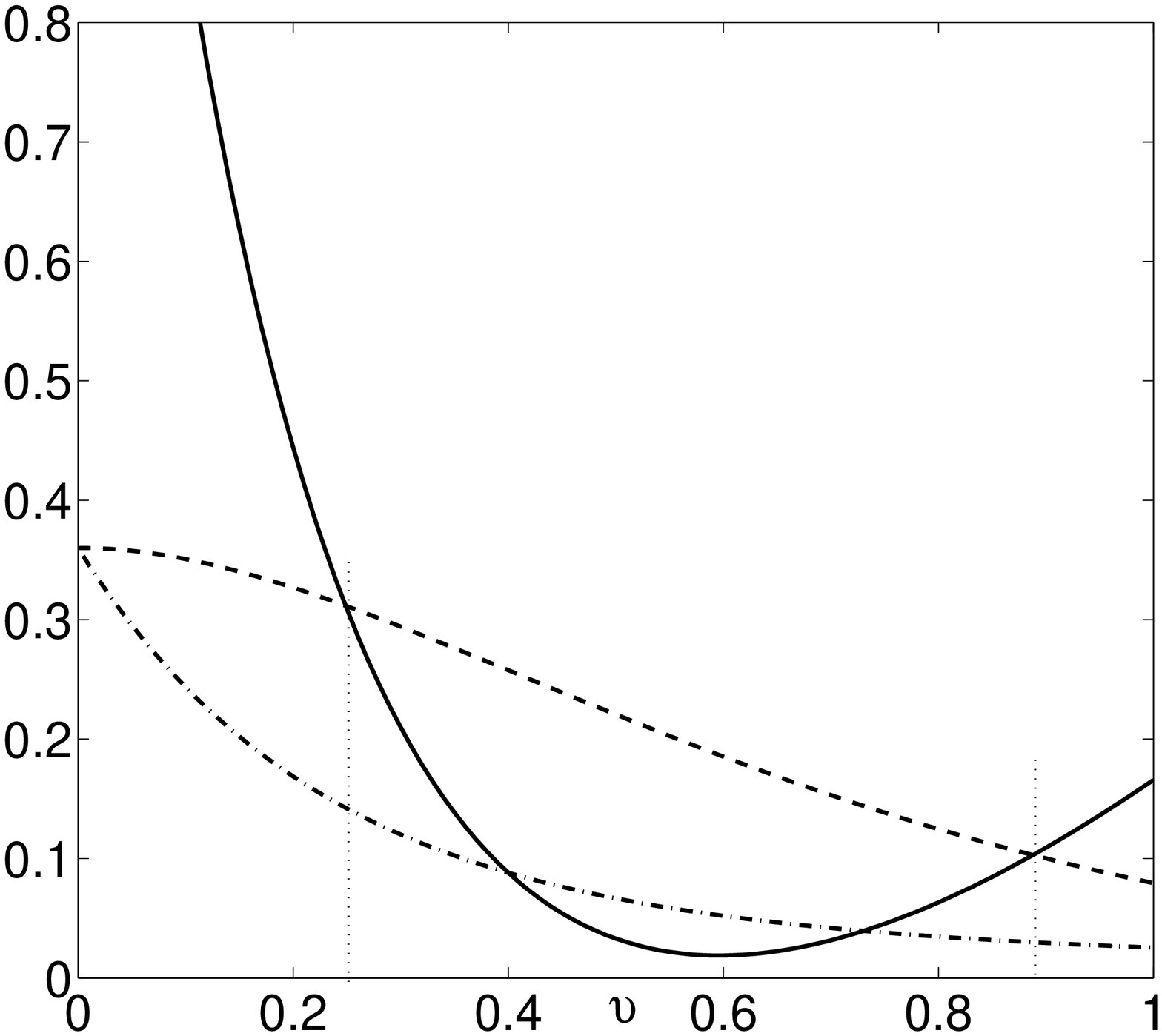,width=0.85\columnwidth}}
\vspace*{13pt}
\fcaption{\label{6}(Dimensionless) The evolution of eigenvalues of matrix $A$. It
should be noted that each type of line corresponds to one separate
eigenvalue of $A$. The crossing points of the larger two 
eigenvalues correspond to the critical points of sudden change shown in
Fig. 5. }
\end{figure}
information theoretic quantum discord by looking for the crossing points of $%
s_{i}$, which is given in the inset of Fig. 5. However, according to our
calculation as well as the illustration in Fig. 5, one can easily find that
the critical points of sudden change of geometric and information theoretic
quantum discords are not consistent. It is obvious that there exist two
critical points for geometric quantum discord, but only one critical point
exists for information theoretic quantum discord. In addition, one can
also find that the sudden changes given by different quantum correlation
measure do not happen at the same time. From the previous examples, we can safely say
that the sudden change of quantum correlation, like the frozen quantum
correlation, strongly depends on the choice of quantum correlation measure.

\section{Conclusion and discussions}

We have introduced a definition of sudden change of quantum correlations,
based on which we present a simple witness on the sudden change in terms of
the geometric quantum discord. It is shown that there is only one kind of
way to leading to sudden change, that is, the crossing of the two larger
eigenvalues of matrix $A$. As applications, we demonstrate the sudden
changes of quantum correlation by considering some quantum systems under
various decoherence processes. One can find out any critical points of
sudden changes using our witness, even though the sudden changes could not
be so obvious in the graphical representation. As comparisons, we
simultaneously consider the information theoretic quantum discord. It is
interesting that, the sudden change will not be consistent if we choose
different quantum correlation measures. This implies that sudden change,
like frozen quantum correlation but unlike the sudden death of quantum
entanglement, strongly depend on the choice of quantum correlation measure.
In other words, sudden change of quantum correlation should not be the
property of quantum state , but that of the quantum correlation measure
subject to some states. From a different angle, it is interesting that
different measure of entanglement may give different ordering of two
bipartite states, this only means that those two states are uncomparable,
i.e., they cannot be converted to each other by local operations and
classical communication. However, considering different measure of quantum
correlation, although they, beyond quantum entanglement, show many
inconsistencies, we can not obtain any clear information on the conversion
between quantum states similar to that of quantum entanglement.

\section{Acknowledgements}

This work was supported by the National Natural Science Foundation of China,
under Grant No. 11175033 and `973' program No. 2010CB922904 and the
Fundamental Research Funds of the Central Universities, under Grant No.
DUT12LK42.

\end{document}